\documentclass[iop]{emulateapj}

\begin{document}
\title{Resolved CO gas interior to the dust rings of the HD 141569 disk}
\author{
Kevin M. Flaherty\altaffilmark{1},
A. Meredith Hughes\altaffilmark{1},
Sean M. Andrews\altaffilmark{2},
Chunhua Qi\altaffilmark{2},
David J. Wilner\altaffilmark{2},
Aaron C. Boley\altaffilmark{3},
Jacob A. White\altaffilmark{3},
Will Harney\altaffilmark{4},
Julia Zachary\altaffilmark{1}
}
\altaffiltext{1}{Van Vleck Observatory, Astronomy Department, Wesleyan University, 96 Foss Hill Drive, Middletown, CT 06459}
\altaffiltext{2}{Harvard-Smithsonian Center for Astrophysics, 60 Garden Street, Cambridge, MA 02138}
\altaffiltext{3}{Department of Physics and Astronomy, University of British Columbia, Vancouver BC, Canada}
\altaffiltext{4}{Department of Physics and Astronomy, Union College, Schenectady, NY}

\begin{abstract}
The disk around HD 141569 is one of a handful of systems whose weak infrared emission is consistent with a debris disk, but still has a significant reservoir of gas. Here we report spatially resolved mm observations of the CO(3-2) and CO(1-0) emission as seen with the SMA and CARMA. We find that the excitation temperature for CO is lower than expected from cospatial blackbody grains, similar to previous observations of analogous systems, and derive a gas mass that lies between that of gas-rich primordial disks and gas-poor debris disks. The data also indicate a large inner hole in the CO gas distribution and an outer radius that lies interior to the outer scattered light rings. This spatial distribution, with the dust rings just outside the gaseous disk, is consistent with the expected interactions between gas and dust in an optically thin disk. This indicates that gas can have a significant effect on the location of the dust within debris disks.

\end{abstract}

\section{Introduction}

Debris disks are characterized by weak scattered light and thermal emission from small dust grains encircling the central star. Most observations focus on this dust emission \citep[and references therein]{wya08,kri10,mat14} but the presence of gas can substantially influence the observed dust structures \citep{tak01,lyr13}. While surveys have found that most debris disks have a negligible gas mass \citep{pas06}, some systems possess significant amounts of gas in addition to their debris-like dust distributions. Radio surveys revealed that HD 21997, a 30 Myr old A star, is surrounded by a broad disk of CO with a mass of M$_{\rm CO}$$\sim$5$\times 10^{-2}$ M$_{\earth}$ \citep{kos13} while 49 Ceti also has a large ring of CO emission, with a mass estimated at M$_{\rm CO}$$\sim$10$^{-3}$ M$_{\earth}$ \citep{hug08}. ALMA observations of $\beta$ Pic find CO gas confined to a belt, with clear asymmetries, and a mass of only $2\times10^{-5}$ M$_{\oplus}$ \citep{den14}. Recently HD 131835 was discovered to have CO emission consistent with M$_{\rm CO}$$\sim4\times$10$^{-4}$ M$_{\earth}$ \citep{moo15}. In addition to these radio observations, optical surveys have revealed gas in the the debris disk systems HR 10 \citep{lan90}, HD 32297 \citep{red07}, 49 Ceti and $\iota$ Cyg \citep{mon12}, HD 2160, HD 110411, HD 145964, HD 183324 \citep{wel13} and HD 17255 \citep{kie14}. $\beta$ Pic has been extensively studied using FUV/optical gas tracers \citep[e.g.][]{pet99,rob00,bra04,rob06} revealing a variable gas component due to the evaporation of comet-like bodies as they approach the central star. Far-infrared gas emission lines have also been detected by Herschel around HD172555 \citep{riv10} and HD 32297 \citep{don13}. The gas revealed by these observations may be left over from the primordial disk, which is expected to dissipate after a few million years \citep{fed10}, or created in a second generation process, such as the evaporation of infalling comets \citep{beu07}, or the collisional release of gas through sublimation \citep{zuc12} or vaporization \citep{cze07}.




HD 141569, a B9.5, 5Myr old star, located 108pc away \citep{mer04}, hosts one of the first debris disks discovered to have a significant gas reservoir \citep{zuc95}. While still fairly young, this system does not exhibit a strong infrared excess; instead its dust emission is more characteristic of an optically thin debris disk with L$_{\rm IR}$/L$_*$$\sim$8$\times10^{-3}$ \citep{syl96}. Scattered light images find small dust grains confined to two narrow rings at 240 and 400 au \citep{bil15}, with a gap between them \citep{wei99,aug99} and evidence for spiral arms in the outermost ring \citep{cla03}. Interior to these rings there is very little scattered light emission from small dust grains, although mid-infrared thermal dust emission has been detected down to tens of au \citep{fis00,mar02,moe10}. The weak dust emission and ring features are typical of a debris disk, but the strong CO emission, indicating a substantial gas mass, is not \citep{zuc95}. The presence of a gas reservoir has implications for the origin and evolution of the dust within this system. Cospatial gas and dust can lead to sharp rings without the need for shepherding planets \citep{lyr13}, while collisions with gas as dust is pushed outwards by radiation pressure can lead to a ring at the outer edge of the gaseous disk \citep{tak01}.

Here we present radio interferometric observations of the CO(3-2) and CO(1-0) emission at spatial resolutions of $\sim1\farcs5$ and $\sim4\farcs5$ respectively. The high spatial resolution of these observations allow us to measure the location of the gas, as well as its overall temperature and density structure. In Section 2 we present the observations, and in Section 3 we present our model and fitting procedure. In Section 4 we present our results, in particular the finding that the gas has a low excitation temperature and is confined to within the scattered light rings, while in Section 5 we discuss the implications of these results for debris disk structure and evolution.

\section{Observations and Results}
We observed HD 141569 with the Submillimeter Array (SMA) over the course of two nights in 2012. On 28 April 2012 we utilized the array's compact configuration, including six of the array's eight antennas spread across projected baseline lengths ranging from 12 to 86\,m. The weather was excellent, with the 225\,GHz atmospheric opacity holding steady between 0.03 and 0.05 throughout the night. Both Titan and Neptune were used as flux calibrators, allowing us to derive a flux density of 1.74\,Jy for the quasar J1512-090, which was used as the primary gain calibrator. Observations of the source were interleaved with brief observations of both J1512-090 and J1549+026, which was included to test the atmospheric gain transfer to the source, over a series of 13.5-minute loops. Passband calibration using the quasar 3C279 (derived flux 13.7\,Jy) occured for one hour at the start of the track. On 10 Feb 2012 we observed HD 141569 with the SMA in its extended configuration, including seven of the array's eight antennas spanning projected baseline lengths of 22 to 252\,m. The weather was good, with the 225\,GHz opacity steady at 0.05 for most of the night, although puncutated by a brief period with values as high as 0.12 for approximately two hours in the middle of the 10-hour track. We used the same calibrators and observing strategy as in April and derived a flux of 1.47\,Jy for the quasar J1512-090 using only Titan as the flux calibrator. All SMA and CARMA flux measurements are subject to a 20\% systematic flux uncertainty that results from uncertainties in the flux models of the solar system objects used to derive the flux scale. 

For both nights the receivers were tuned to the CO(3-2) rest frequency of 345.79899\,GHz. The CO(3-2) line was observed in the center of the lower 2\,GHz of the 4 GHz-wide upper sideband, with a spectral resolution of 256 channels across the 104\,MHz correlator ``chunk'': the remainder of the effective 4\,GHz bandwidth was observed at a lower spectral resolution of 64 channels per 104\,MHz chunk and subsequently averaged together to form a single continuum channel with an effective bandwidth of approximately 8\,GHz. 

Observations of the HD 141569 system with CARMA took place over the course of two nights in April 2012. On 17 April 2012 we observed the source for four hours using CARMA's D configuration with maximum baseline length 146\,m. The weather was very good for 3\,mm observations at the Cedar Flat site, with the 230\,GHz opacity averaging 0.19 over the course of the night and the sky RMS averaging 219\,$\mu$m. We continued D configuration observations for an additional four hours on the night of 19 April 2012. The weather remained stable, with the 230\,GHz opacity averaging 0.23 and the sky RMS averaging 146\,$\mu$m. Mars was used as the primary flux calibrator for both nights. We derived a flux density for the primary gain calibrator, the quasar J1549+026, of 1.48\,Jy on 17 April and 1.42\,Jy on 19 April. The quasar 3C279 was observed for 15 minutes at the start of the 17 April track and used to calibrate the bandpass; since no standard passband calibrator was observed during the 19 April track, we used the quasar J1512-090 as the passband calibrator and obtained satisfactory solutions. 

For both nights, the first local oscillator frequency was set to 108.0314\,GHz. Two spectral windows were tuned respectively to CO(1-0) and $^{13}$CO(1-0) rest frequencies of 115.27\,GHz and 110.201\,GHz. These spectral windows were observed at a spectral resolution of 0.488\,MHz, with the line placed in the upper sideband. The remaining six spectral windows were each set the maximum available bandwidth of 0.468\,GHz to maximize continuum sensitivity, resulting in an aggregate bandwidth of 5.6\,GHz across the lower and upper sidebands.

\begin{deluxetable*}{ccccc}
\tablewidth{0pt}
\tablecaption{Emission Properties\label{base_data}}
\tablehead{\colhead{} & \colhead{CO(3-2)} & \colhead{CO(1-0)} & \colhead{870$\micron$} & \colhead{2.8mm}}
\startdata
Channel width (km s$^{-1}$) & 0.35 & 1.3 & - & -\\
Beam size (FWHM)\tablenotemark{a} & $1\farcs51\times1\farcs02$ & $4\farcs8\times4\farcs0$ & $1\farcs66\times1\farcs16$ & $5\farcs1\times4\farcs2$\\
P.A. & -83.0$\degr$ & 28.6$\degr$ & -82.2$\degr$ & 29.1$\degr$\\
rms (mJy beam$^{-1}$) & 81 & 31 & 0.7 & 0.3\\
Integrated intensity & 15.8 $\pm$ 0.6\tablenotemark{b} Jy km s$^{-1}$ & 1.6 $\pm$ 0.2\tablenotemark{b} Jy km s$^{-1}$ & 8.2$\pm$2.4\tablenotemark{c} mJy & 0.78$\pm$0.3\tablenotemark{d} mJy\\
Peak Flux (Jy beam$^{-1}$) & 1.39 & 0.31 & 4.1$\times10^{-3}$ & 1.1$\times10^{-3}$\\
\enddata
\tablenotetext{a}{Derived assuming natural weighting} 
\tablenotetext{b}{Derived by integrating the moment 0 map, within the 3$\sigma$ contour. Uncertainty is the 1$\sigma$ noise on the integrated intensity.}
\tablenotetext{c}{Derived by fitting an elliptical gaussian to the short-baseline visibilities (uv$<$50k$\lambda$) using the MIRIAD task {\it uvfit}. Uncertainty is the 1$\sigma$ noise on the integrated intensity.}
\tablenotetext{d}{Derived by fitting a point source to the visibilities using the MIRIAD task {\it uvfit}. Uncertainty is the 1$\sigma$ noise on the integrated intensity.}
\end{deluxetable*}

\begin{figure*}
\center
\includegraphics[scale=.25]{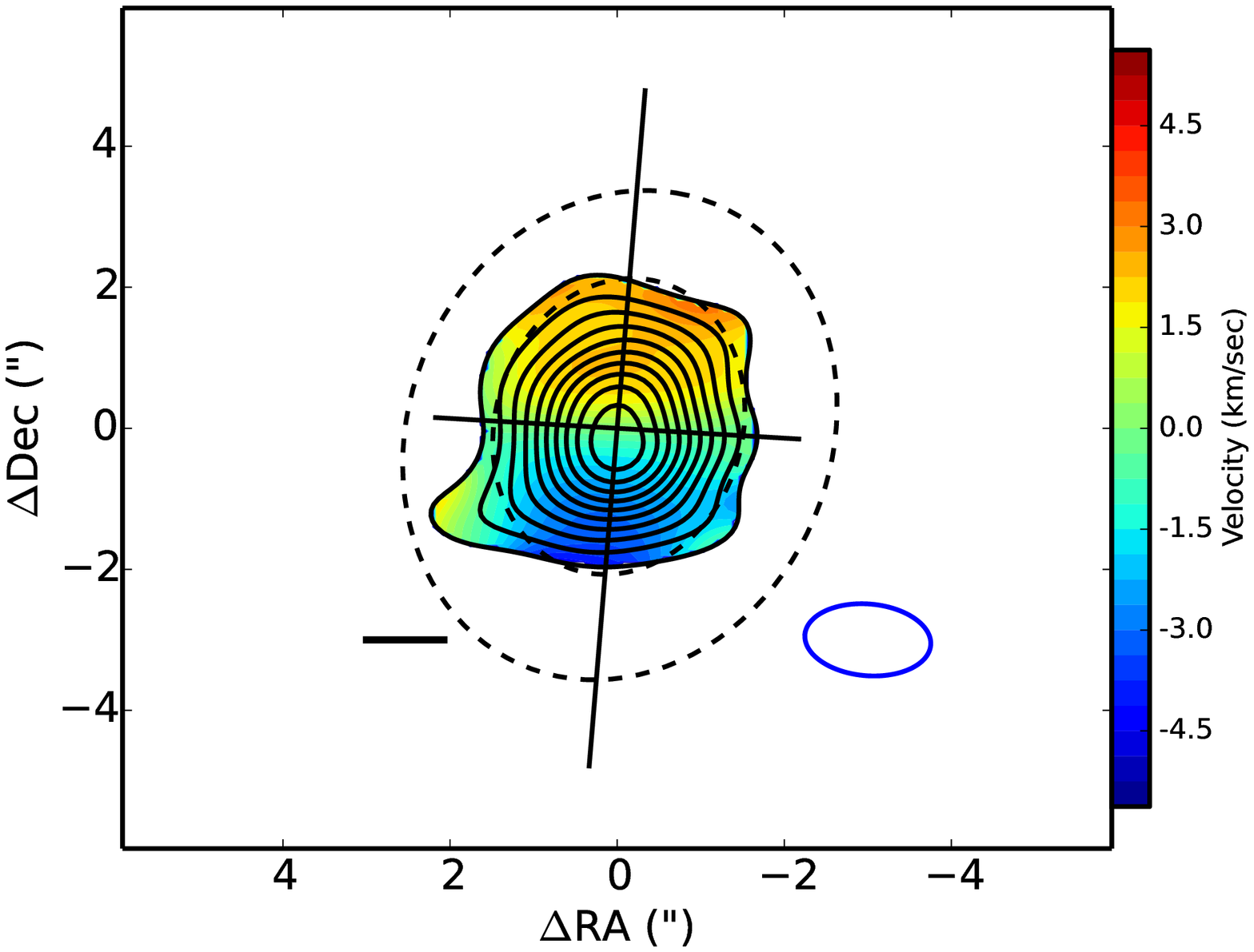}
\includegraphics[scale=.25]{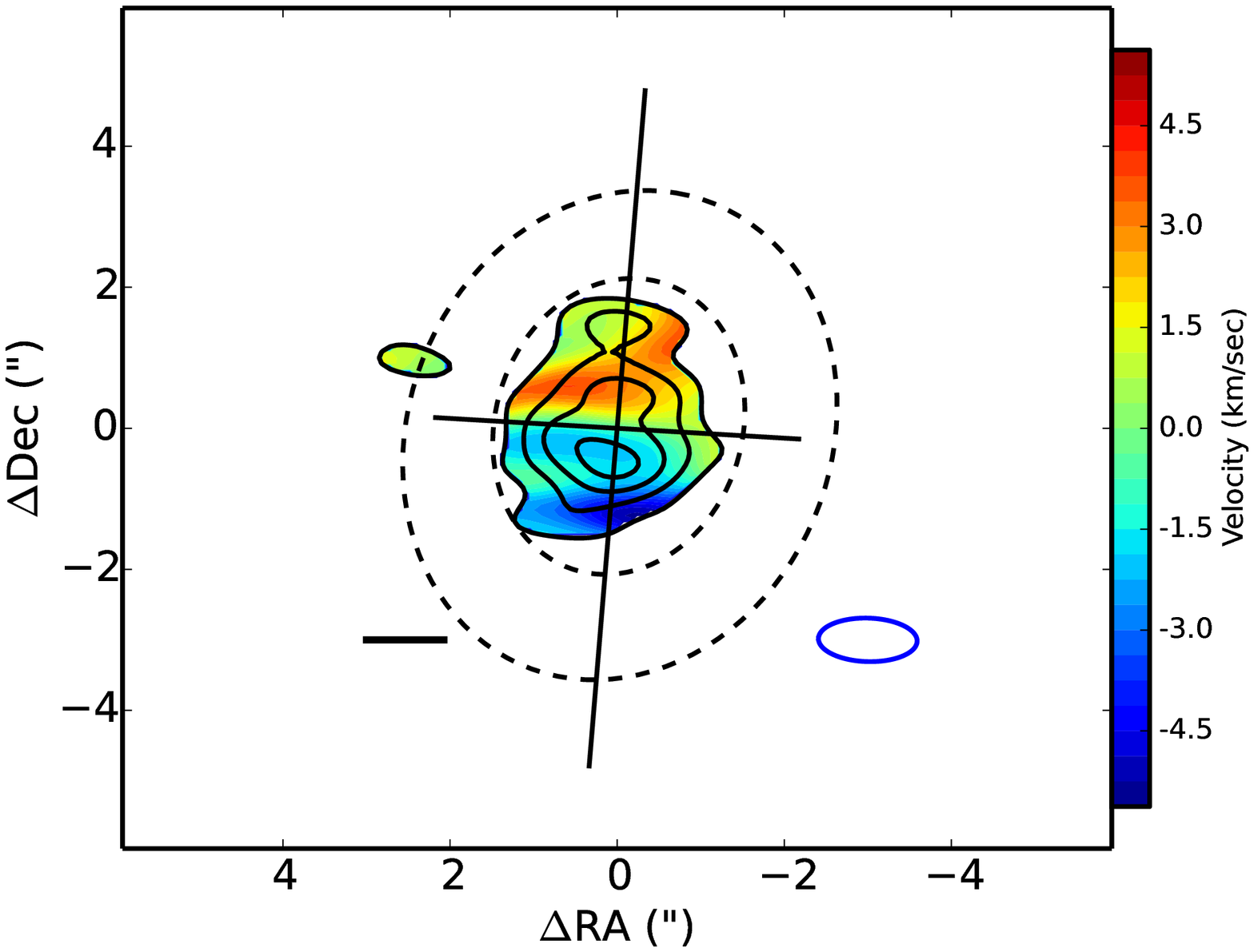}
\includegraphics[scale=.25]{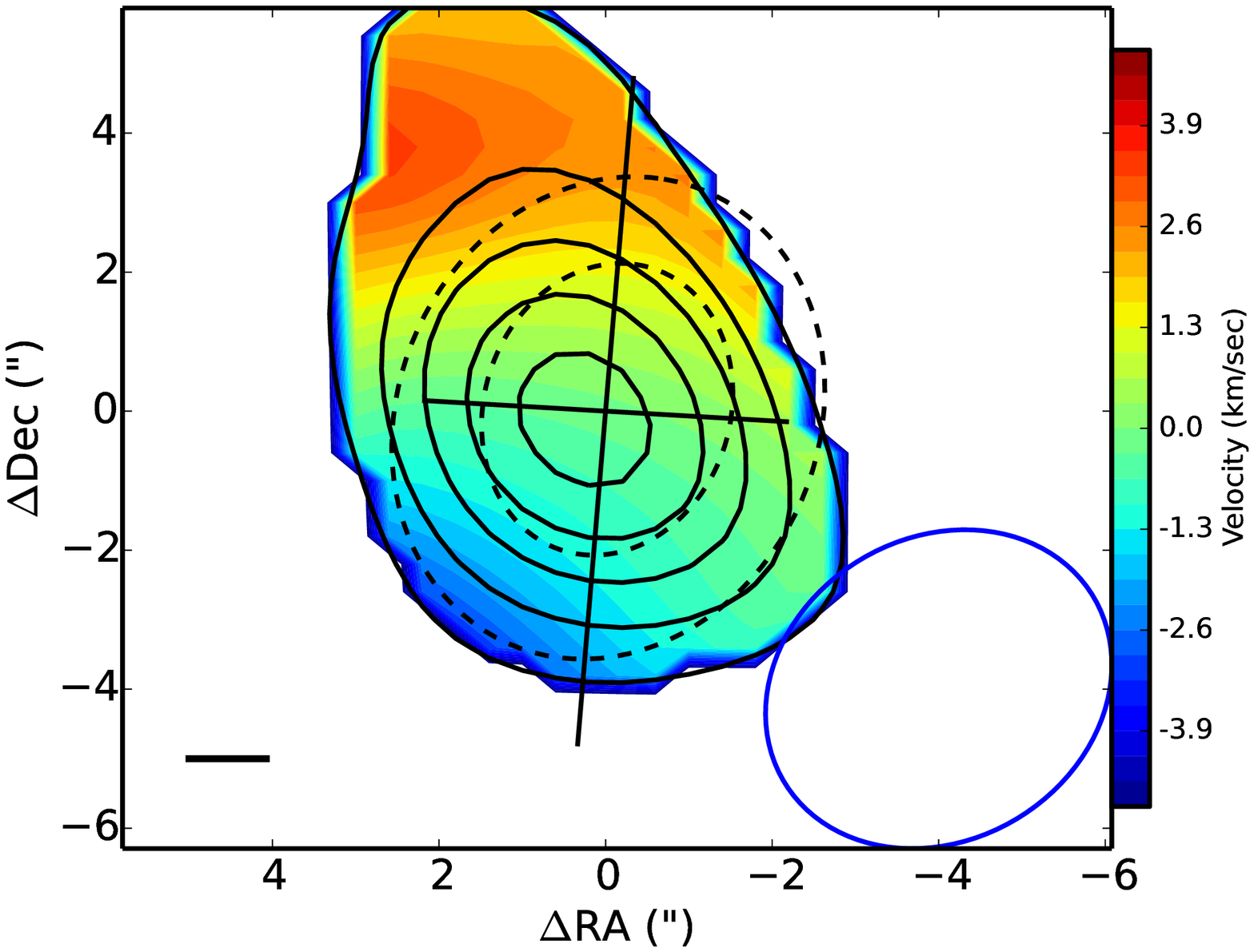}
\caption{(Left) CO(3-2) moments zero (total intensity, black contours) and one (intensity-weighted velocity, filled contours). Intensity contours are in units of 3$\sigma$, 5$\sigma$, 7$\sigma$,... ($\sigma$=0.24Jy/beam). The two dashed circles mark the locations of the two rings seen in scattered light\citep{bil15}. The CO gas is not cospatial with the small dust grains, but is instead located within the inner disk. (Middle) CO(3-2) moment zero and ones maps with uniform weighting. While the noise is higher ($\sigma$=0.32Jy/beam) the higher spatial resolution highlights the double-peaked structure, which is indicative of an inner hole in the CO emission. (Right) CO(1-0) moments zero (total intensity, black contours) and one (intensity-weighted velocity, filled contours). Intensity contours are in units of 3$\sigma$, 5$\sigma$, 7$\sigma$,... ($\sigma$=0.12Jy/beam). In all panels the black horizontal bar marks a length scale of 100 au at a distance of 108 pc, while the circle in the lower right indicates the beam size.\label{moments}}
\end{figure*}

Both lines are clearly detected at high signal-to-noise in our data (Table~\ref{base_data}). The moment maps for CO(3-2) and CO(1-0), derived using the MIRIAD \citep{sau95} task {\it moment} and shown in Figure~\ref{moments}, demonstrate that the spatially resolved CO(3-2) emission is confined to within the scattered light rings. A uniform disk fit to the CO(3-2) emission, using the MIRIAD task {\it imfit}, derivess an outer radius of 290 au for the CO gas, similar to the location of the inner dust ring \citep{bil15}. This is similar to the outer radius derived from the CO line profile in single-dish observations \citep{den05}. Our integrated flux measurements are similar to those of previous CO(3-2) observations of the HD 141569 disk \citep{zuc95,den05}. The improved spatial resolution of uniform weighting of the CO(3-2) data ($1\farcs2\times0\farcs6$ vs $1\farcs5\times1\farcs0$ for natural weighting) reveals a double-peaked structure indicative of an inner hole in the CO gas (Figure~\ref{moments}). The two peaks in the moment 0 map are separated by roughly one beam FWHM, which implies a radius for the inner hole of $<$50 au, consistent with the 11 au inner hole inferred using spectro-astrometric measurements of mid-infrared CO emission. 

With clear detections of two spectral lines, we have additional information on the temperature of the gas. Their flux ratio of 10$\pm$1 implies an excitation temperature of 17$\pm$2 K, assuming both are optically thin, which indicates that the gas is very cold. The CO(1-0) flux implies a CO mass of M$_{\rm CO}$ = 1.05$\pm0.06\times10^{-8}$ M$_{\odot}$, assuming a temperature derived from the line ratio, which is similar to 49 Ceti, HD 21997 and HD 131835 \citep{hug08,kos13,moo15}. This corresponds to M$_{\rm gas}$=1.05$\pm0.06\times10^{-4}$ M$_{\odot}$=0.11$\pm0.01$ M$_{\rm jup}$, assuming CO/H$_2$=10$^{-4}$, which is not enough to create a Jupiter-sized gas giant, but is larger than typically found among debris disks \citep{pas06}. This mass falls within the range inferred by \citet{zuc95} from single-dish CO observations (Table~\ref{mass_measurements}). These simple estimates paint a picture of a radially extended cold gas disk whose mass is still significant, although it is diminished relative to younger systems.



In addition to the gas, we also detect continuum emission from the dust at 870$\mu$m, with peak S/N$\sim$6, and marginally detect the dust at 2.8mm, with peak S/N$\sim$4. Our observed radio emission is unresolved in our $1\farcs5\times1\farcs0$ beam, suggesting that the mm-sized dust is concentrated in the inner disk. Using the MIRIAD task {\it uvfit} we fit a Gaussian to the short baseline 870$\micron$ visibilities and measure a total flux of 8.2$\times10^{-3}$ Jy (Table~\ref{base_data}). Previous single-dish measurements of the continuum measure fluxes of 10.9$\pm$1.3 mJy at 850$\micron$ \citep{san11} and 12.6$\pm$4.6mJy at 870$\micron$ \citep{nil10}. Our 870$\micron$ continnum flux of 8.2$\pm$2.4 mJy is consistent with, although slightly smaller than, these measurements likely due to missing short-baseline flux. Assuming the dust is optically thin, our measured flux can be converted to a mass of mm-sized dust by
\begin{equation}
M_{\rm dust} = \frac{F_{\nu}d^2}{\kappa_{\nu} B_{\nu}(T_{\rm d})},
\end{equation}
where $F_{\nu}$ is the flux density, $d$ is the distance, $\kappa_{\nu}$ is the mass opacity and $B_{\nu}(T_d)$ is the planck function evaluated at a dust temperature of $T_d$ \citep[e.g.][]{and13}. This flux is only sensitive to the small dust grains; in a system with collisionally generated dust grains a substantial amount of mass is locked in meter to km-sized planetesimals that do not emit efficiently at these wavelengths and any mass inferred from the radio emission underestimates the total mass. With this caveat in mind, we derive a mass in small dust grains of M$_{\rm dust}$=1.8$\times10^{-6}$ M$_{\odot}$=0.6 M$_{\oplus}$ assuming an opacity (1.5 cm$^2$ g$^{-1}$ more appropriate for $\sim$mm-sized dust grains in protoplanetary disks, and the temperature (51 K) of a blackbody dust grain at 150 AU from the central star. Assuming a gas to dust mass ratio of 100, this implies M$_{\rm gas}$=1.8$\times10^{-4}$ M$_{\odot}$, which is similar to the gas mass derived from the CO(1-0) emission, although both are subject to large systematic uncertainties due to the assumptions about dust/gas temperature, the dust opacity, the gas to dust mass ratio, the CO/H$_2$ conversion factor, and the ability of continuum emission to trace the full range of grain sizes.
Our dust-based disk mass estimate is larger than derived previously (Table~\ref{mass_measurements}), \citet{san11} find M$_{\rm gas}$=9$\times10^{-5}$ M$_{\odot}$ using single-dish 850$\micron$ flux measurement while \citet{li03} derive M$_{\rm gas}$=1.1$\times10^{-5}$ M$_{\odot}$ from SED modeling. The discrepancy could be due to differences in the dust temperature, SED modeling of infrared emission finds temperatures of 70-200 K \citep{li03} while the single dish measurements assume 40 K \citep{san11}, or differences in the assumed dust opacity. Despite the differences in the various studies, and the limited ability of sub-mm emission to trace the total dust mass, it is clear the HD 141569 is surrounded a low mass disk of gas and dust.

\begin{deluxetable*}{ccc}
\tablewidth{0pt}
\tablecaption{Disk Mass Measurements\label{mass_measurements}}
\tablehead{\colhead{M$_{\rm disk}$(M$_{\odot}$)} & \colhead{Method} & \colhead{Reference}}
\startdata
6-140$\times10^{-5}$ & Single-dish CO\tablenotemark{a} & \citet{zuc95}\\
1.1$\times10^{-5}$ & SED modeling\tablenotemark{b} & \citet{li03}\\
2.4$\times10^{-4}$ & Radiative transfer modelling\tablenotemark{c} & \citet{jon07}\\
9$\times10^{-5}$ & dust continuum\tablenotemark{d} & \citet{san11}\\
2.5-5$\times10^{-4}$ & Radiative transfer modelling\tablenotemark{c} & \citet{thi14}\\
\hline
1.8$\times10^{-4}$ & dust continuum\tablenotemark{d} & this paper\\
1.05$\times10^{-4}$ & CO(1-0)\tablenotemark{a} & this paper\\
\hline
3.8$^{+15}_{-2.8}\times10^{-5}$ & MCMC modeling & this paper\\
\enddata
\tablecomments{Estimates of the gas mass of the disk around HD 141569A}
\tablenotetext{a}{Gas mass is derived assuming the observed line is optically thin and tracers the entire H$_2$ mass with a constant CO/H$_2$ abundance.}
\tablenotetext{b}{Gas mass is derived from fitting the full SED with a dust model, and extrapolating to the gas mass}
\tablenotetext{c}{Gas mass is derived using radiative transfer model fitting to multiple gas and dust emission diagnostics}
\tablenotetext{d}{Gas mass is derived from the mm dust continuum emission, assuming a gas to dust mass ratio}
\end{deluxetable*}

\section{Modeling the Gas Emission}
While a preliminary analysis of the CO emission indicates a significant mass distributed inside the scattered light rings, detailed modeling is needed to determine the uncertainties on this structure. To characterize the emission in greater detail we employ a simple parametric model and derive the posterior distribution functions (PDFs) for each model parameter, using a Markov-Chain Monte-Carlo technique. Our model of the temperature structure is a radial power law with an isothermal vertical profile. 
\begin{equation}
T_{\rm gas}(r) = T_{0}\left(\frac{r}{150 \rm \ au}\right)^{q}.
\end{equation}

We assume that the surface density is a power law, with index $\gamma$, between $R_{\rm in}$ and $R_{\rm out}$.
\begin{equation}
\Sigma_{\rm gas}(r) = \frac{M_{\rm gas}(2-\gamma)}{2\pi(R_{\rm out}^{2-\gamma}-R_{\rm in}^{2-\gamma})}r^{-\gamma}.
\end{equation}
In a vertically isothermal disk the density profile is specified by $\rho(r,z) = \rho(r,z=0)\exp(-(z/H)^2)$ where $H$ is the pressure scale height ($H=\sqrt{2}c_s$/$\Omega$) and $\rho(r,z=0)=\Sigma_{\rm gas}(r)/(\sqrt{\pi} H)$. The isothermal sound speed is given by $c_s^2=k_BT_{\rm gas}/\mu m_h$, and $\Omega$ is the Keplerian rotation frequency. The disk is assumed to be made of 80\%\ molecular hydrogen by mass with $\mu=2.37$. We spread CO throughout the disk, with a spatially constant CO to H$_2$ abundance of 10$^{-4}$. While we assume that CO traces the entire gas disk, this may not be the case. Freeze-out onto dust grains and photo-dissociation by high energy photons can remove CO from the system. While the dust temperature is high enough that freeze-out is unlikely \citep{li03}, photodissocation at the inner and outer edges of the disk may still be a factor. \citet{jon07} find that the gas mass of the disk around HD 141569 is in a regime where self-shielding begins to set in, resulting in large changes in CO structure in result to small variations in CO mass. As discussed below there is evidence of gas within our derived inner radius based on measurements of ongoing accretion, while the morphology of the CO relative to the scattered light rings suggests that little gas extends beyond the outermost CO edge.

The flux is calculated along sight lines through the disk according to:
\begin{equation}
I_{\nu} = \int^{\infty}_0S_{\nu}(s)\exp[-\tau_{\nu}(s)]K_{\nu}(s)ds
\end{equation}
where $s$ is the linear coordinate along the line of sight increasing outward from the observer. The optical depth is $\tau_{\nu}(s)=\int_0^sK_{\nu}(s')ds'$ where $K_{\nu}(s)$ is the absorption coefficient and $S_{\nu}(s)$ is the source function. The line is assumed to be a Gaussian with a width given by thermal motion, with zero turbulent motion. We also assume the line is in Local Thermodynamic Equilibrium (LTE), an assumption that is consistent with the observed disk properties, as we confirm below. The disk structure and radiative transfer calculation is carried out using code originally developed in \citet{ros12,ros13} and \citet{fla15} with modifications for our simplified disk structure. The central star is assumed to lie at a distance of 108 pc \citep{mer04} with a position angle of 356$^{\circ}$ \citep{wei00}. Unless otherwise specified, we also assume a stellar mass of 2 M$_{\odot}$ \citep{mer04}. The disk emission in the SMA data is offset by [-0.2'',-0.2''] from the phase center due to the proper motion of the system between the J2000 epoch coordinates and when the data were taken; this offset is included in our model. No spatial offset is applied to the CARMA data. 

We use the affine-invariant Markov-Chain Monte-Carlo routine EMCEE \citep{for13} to sample the posterior probability distribution. Our initial sampling of the posterior distribution functions involves a smaller number of walkers run for a substantial length of time (e.g. 80 walkers over 1000 steps), followed by a more detailed sampling of the PDFs using more walkers, whose starting positions are spread throughout the 3$\sigma$ ranges defined from the initial run, for fewer steps (e.g. 200 walkers over 300 steps). Initial steps corresponding to burn-in are removed from the chains. The goodness of fit between the model and observed visibilities is calculated using the chi-squared statistic, accounting for the complex visibility weights. Model visibilities are calculated from the model images using the MIRIAD task {\it uvmodel}. Our fitting uses both the CO(3-2) and CO(1-0) lines simultaneously by summing the $\chi^2$ from each line. The starting positions of the walkers during the initial sampling are chosen to be evenly distributed throughout parameter space over an area that covers the expected best fit (Table~\ref{init_params}). Priors are placed on the parameters to ensure physically realistic models, including $0<M_{\rm disk}<M_*$, $R_{\rm in}>0$, $R_{\rm out}>R_{\rm in}$, $T_0>0$, and $-3<\gamma<3$. With only 2-3 resolution elements across the radius of the disk we do not have a strong constraint on the radial profile of the emission. For this reason we fix $q=-0.5$ while allowing $\gamma$ to vary. Our final list of variable model parameters is [$M_{\rm gas}$,$R_{\rm out}$,$R_{\rm in}$,$T_0$,$i$,$\gamma$].

\section{The Results}

\begin{figure*}
\center
\includegraphics[scale=.5]{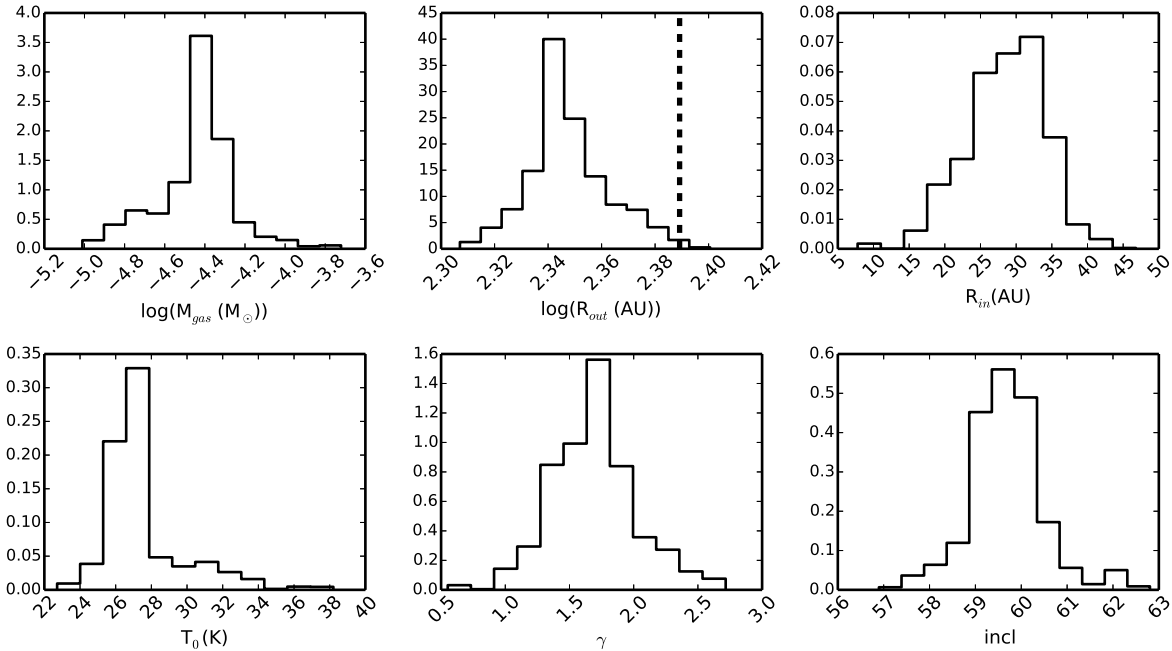}
\caption{Marginalized PDFs for the parameters in our model. The vertical dashed line in the PDF of log($R_{\rm out}$) marks the location of the innermost scattered light dust ring.  \label{pdfs}}
\end{figure*}

\begin{figure}
\center
\includegraphics[scale=.4]{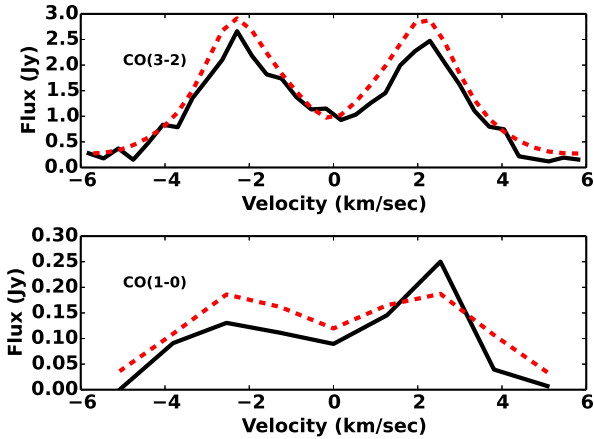}
\caption{Spectra derived from images within a 10'' box for CO(3-2) (top panel) and CO(1-0) (bottom panel). Data are shown with the solid black lines, while model spectra are shown with red dashed lines. \label{imspec}}
\end{figure}

\begin{figure*}
\includegraphics[scale=.25]{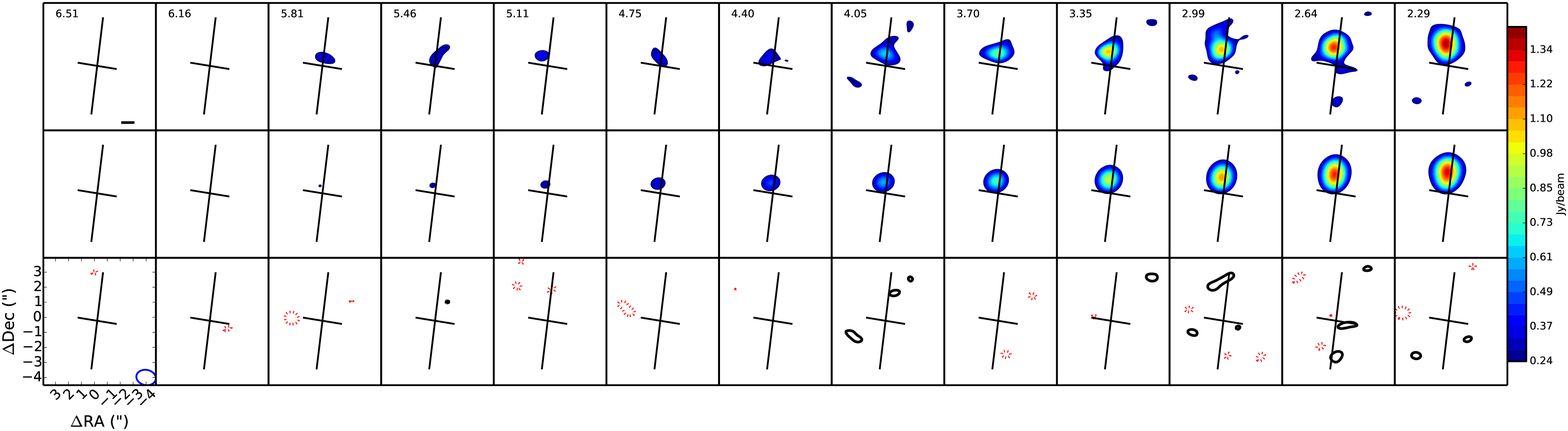}
\includegraphics[scale=.25]{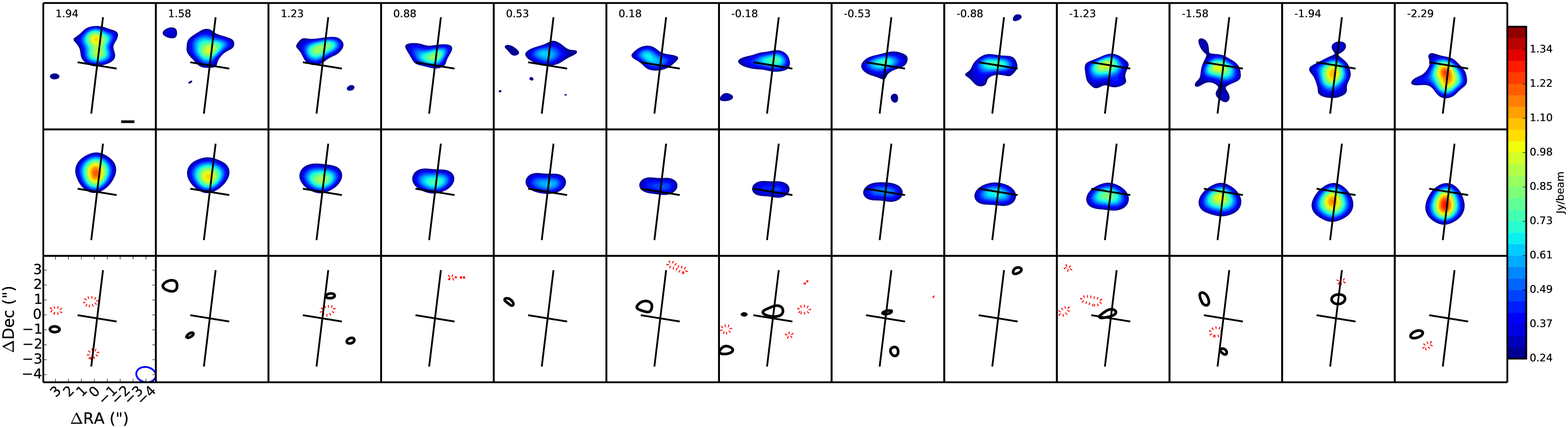}
\includegraphics[scale=.25]{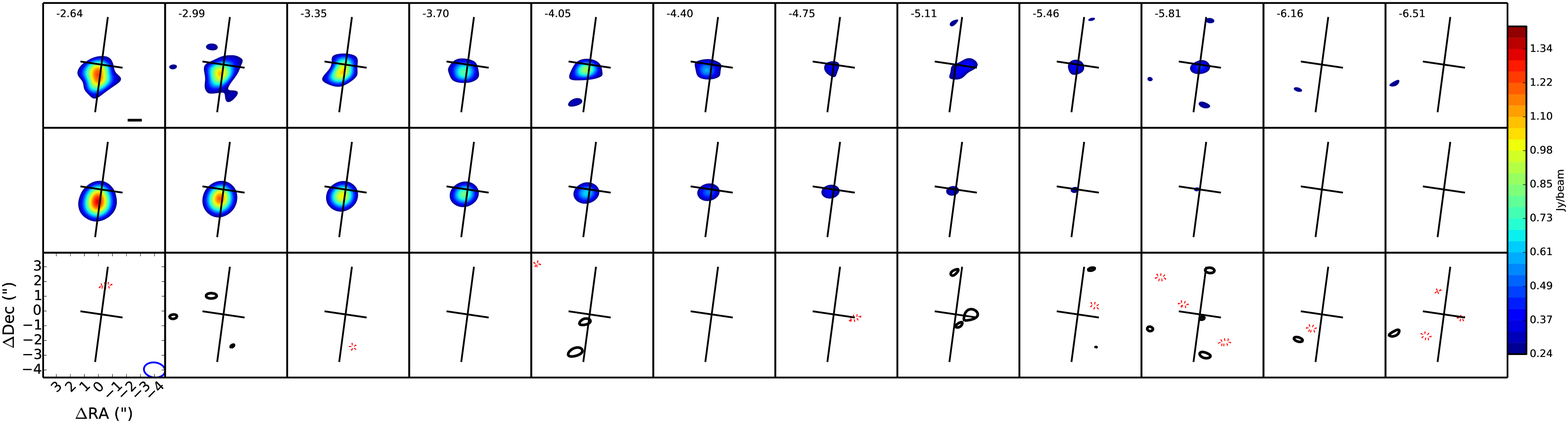}
\caption{Data (top row), model (middle row) and residuals (bottom row) for the CO(3-2) data. Data and model start at 3$\sigma$ ($\sigma=0.08$Jy/beam), while for the imaged residuals the contours are at 3$\sigma$,5$\sigma$,7$\sigma$,... with black solid contours where data$>$model and red dashed contours where model$>$data. Beam size and a 100 au scale bar are marked in the bottom and top row, respectively, of the first column. The first two and last two channels were not included in the fitting, but are shown here to illustrate the noise properties in signal-free channels.\label{residuals_co32}}
\end{figure*}

\begin{figure*}
\includegraphics[scale=.25]{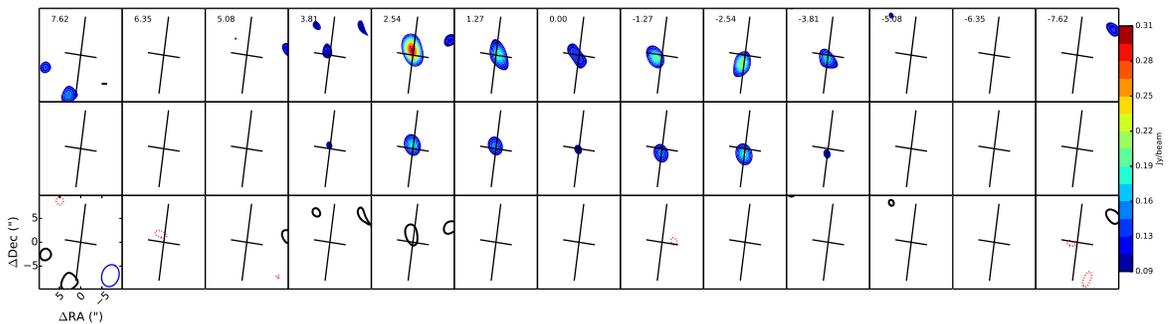}
\caption{Similar to Figure~\ref{residuals_co32}, but with CO(1-0). As with CO(3-2), the first and last two channels were not used in the fitting. Outside of the 2.54 km sec$^{-1}$ channel, there are only minor residuals. \label{residuals_co10}}
\end{figure*}

The PDFs for each parameter are shown in Figure~\ref{pdfs}, with median values and 3$\sigma$ uncertainties listed in Table~\ref{model_fit}. We find that the data are well fit with CO gas that extends from $R_{\rm in}$=29$^{+14}_{-20}$ au to $R_{\rm out}$=224$^{+21}_{-20}$ au with a total gas mass log($M_{\rm gas}$ (M$_{\odot}$))=-4.4$^{+0.7}_{-0.6}$ and a surface density power law index of $\gamma=$1.7$^{+1.0}_{-1.1}$, along with a temperature profile that has $T_0$=27$^{+11}_{-4}$ K. We can place strong constraints on the disk mass, temperature, outer radius and inclination, while there are substantial uncertainties on the inner radius and $\gamma$. While some constraint on $R_{\rm in}$ is provided by the high-velocity channels of the spectra, the limited spatial resolution prevents tight limits from being placed on both $R_{\rm in}$ and $\gamma$. Figure~\ref{imspec} shows the CO(3-2) and CO(1-0) spectra along with the best fit model drawn from the posterior distributions, which is able to match the strength and shape of both spectra. Residuals, derived by subtracting the model and data visibilities from each other and creating a cleaned image of these differences, are shown in Figures~\ref{residuals_co32} and \ref{residuals_co10}. 

\begin{deluxetable}{c}
\tablewidth{0pc}
\tablecaption{Initial Parameter Space\label{init_params}}
\tablehead{ \colhead{Parameter}}
\startdata
-5 $<$ log(M$_{\rm gas}$(M$_{\odot}$)) $<$ -3 \\
2 $<$ log(R$_{\rm out}$(au)) $<$ 2.6 \\
5 au $<$ R$_{\rm in}$ $<$ 60 au\\
10 K $<$ T$_0$ $<$ 40 K\\
0.5 $<$ $\gamma$ $<$ 2\\
50$\degr$ $<$ incl $<$ 70$\degr$\\
\enddata
\end{deluxetable}

\begin{deluxetable}{cc}
\tablewidth{0pc}
\tablecaption{Model Fit\label{model_fit}}
\tablehead{ \colhead{Parameter} & \colhead{Result}}
\startdata
log(M$_{\rm gas}$(M$_{\odot}$)) & -4.4$^{+0.7}_{-0.6}$\\
log(R$_{\rm out}$(au)) & 2.31$\pm0.04$\\
R$_{\rm in}$ & 29$^{+14}_{-20}$ au\\
T$_0$ & 27$^{+11}_{-4}$ K\\
$\gamma$ & 1.7$^{+1.0}_{-1.1}$\\
incl ($\degr$)& 60$\pm3$\\
$\chi^2_{\rm red}$ & 1.91\tablenotemark{a}\\
\enddata
\tablecomments{Median of the marginalized PDF for each parameter, with 3$\sigma$ uncertainties that are defined by the area under the PDF that contains 99.7\%\ of the samples.}
\tablenotetext{a}{Reduced chi-squared, summing the contribution from each line, for the model defined by the medians of the marginalized PDFs}
\end{deluxetable}

\subsection{Surface Density}

\begin{figure}
\center
\includegraphics[scale=.3]{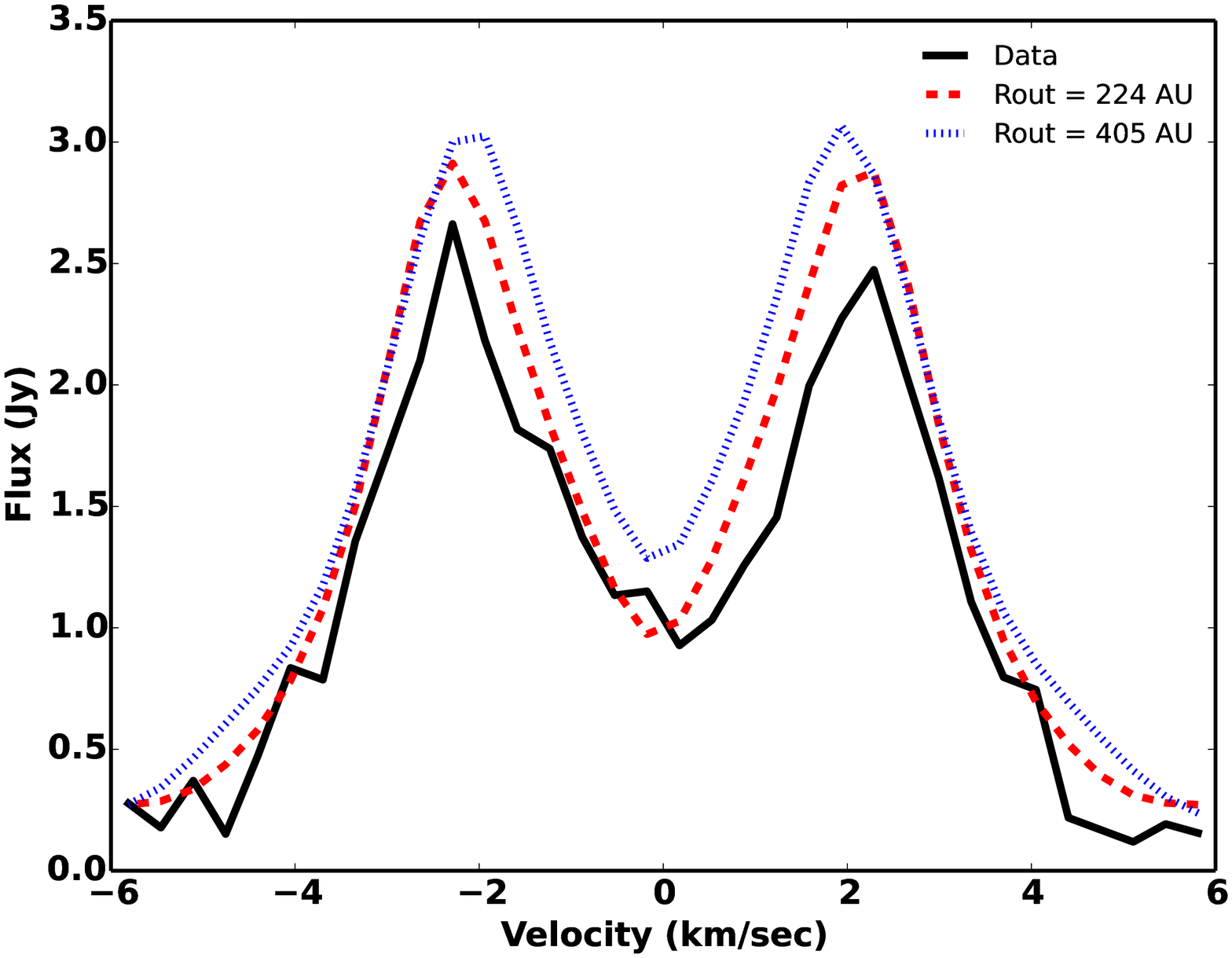}
\caption{CO(3-2) spectra comparing the data (black solid line) to models with various outer radii. When allowed to vary, the outer radius settles at 224$^{+21}_{-20}$ au (red dashed line). If fixed at the location of the outermost scattered light ring (406 au, blue dashed line) the fit is significantly worse. This confirms the lack of a substantial gas mass extending to the outermost dust ring.\label{imspec_Rout}}
\end{figure}

Interestingly we find that the CO gas does not overlap with the scattered light rings. These scattered light rings are at a radius of roughly 245 au and 400 au \citep{wei99,aug99,wei00,cla03,bil15}, while the 3$\sigma$ upper limit on the CO gas outer radius is 245 au. Forcing the models to include gas coincident with the scattered light rings, by extending $R_{\rm out}$ to 405 au, and re-running the MCMC modeling routine, produces a significantly worse fit as the gas is forced to extend further out in the disk (Figure~\ref{imspec_Rout}). The walkers are driven towards very steep $\gamma$, eventually running into our arbitrary upper limit of 3 as these models try to accommodate the large extent of the disk with a very steep density gradient that minimizes the flux from the outer disk. Despite this, they still predict a surface density of $\Sigma\sim4\times10^{-5}$ g cm$^{-2}$ between the two rings, which is large enough to produce detectable emission, although none is seen in the data. The poor quality of these large radius models confirms that there is negligible extended CO gas at the location of the outer rings.

The marginalized PDF of R$_{\rm in}$ indicates that we can constrain the inner radius of the CO. Our result is consistent, within the errors, with spectro-astrometric measurements of the CO v=2-1 line which find $R_{\rm in}$=11$\pm$2 au \citep{got06}. Our ability to constrain the inner radius comes from the spatial and spectral resolution as well as model dependent factors. At uniform weighting, our beam has a FWHM of 0$\farcs$6 along the major axis of the disk, equivalent to a diameter of $\sim$50 au, or a radius of 30 au at the center of the disk, which corresponds to the median of the marginalized PDF. Similarly the line wings probe high velocity gas arising from the center of the disk; a projected velocity of 5 km sec$^{-1}$ corresponds to material at 50 au. Part of our constraint is also model dependent. In the context of the continuous power law surface density and temperature structures that we have chosen, the flux in the central beam can be reduced by increase the inner radius. The consistency between our results and the previous high resolution infrared observations suggests that our result is not strongly influenced by these model dependent effects.


We derive a disk mass of M$_{\rm gas}$=3.8$^{+15}_{-2.8}\times10^{-5}$ M$_{\odot}$=13$^{+50}_{-9}$ M$_{\oplus}$ indicating a small, but still substantial, amount of gas within the disk, consistent with our simple estimates earlier (Table~\ref{mass_measurements}). Our result is smaller than derived from modeling of unresolved emission \citep{thi14,jon07} although these models place a substantial amount of mass in the cold outer disk, which is not present in our resolved observations. The surface density falls off with a power law exponent of 1.7$^{+1.0}_{-1.1}$, similar to the minimum mass solar nebula \citep{wei77} and protoplanetary disks \citep{and09,ise09,and10}, although our result is poorly constrained due to the lack of spatial resolution. 

Our inclination (60$\pm3$) is slightly higher than that derived from scattered light observations (51$\pm$3) \citep{wei99}. This may be due to an underestimate of the stellar mass; White et al. (in prep) find, in their modeling of ALMA CO(3-2) observations, that the velocity profile can be fit with an inclination consistent with the scattered light observations when including a larger stellar mass. We find that fixing the inclination at 51$^{\circ}$ and allowing stellar mass to vary results in a satisfactory fit to the emission profiles with M$_*$=2.15$^{+0.13}_{-0.08}$ M$_{\odot}$, larger than the $M_*=2.00^{+0.18}_{-0.15}$ M$_{\odot}$ derived from analysis of the optical spectrum \citep{mer04}.
While further work is needed to understand the source of this discrepancy, it does not substantially affect our conclusions about the structure of the disk. We find that the disk parameters ($M_{\rm gas}$, $R_{\rm in}$, etc.) fall within the 3$\sigma$ ranges derived from the fiducial model.


\subsection{Excitation Temperature}

An optically thin disk of blackbody dust grains around HD 141569 would have $T_0=51$ K. In the low density environment of a debris disk the gas temperature will not be equal to that of the dust \citep{kam01} and indeed the ratio of the line fluxes, assuming both are optically thin, implies $T=17\pm2$ K. While this simple estimate is a lower limit, since CO(3-2) is not entirely optically thin, our more detailed MCMC modeling finds $T_0$=27$^{+11}_{-4}$ K, which is still significantly lower than the blackbody dust temperature. Fitting the data with models restricted to have $T_0$=51 K finds solutions that underproduce the CO(1-0) flux, consistent with the need to lower the excitation temperature to more heavily populate the J=1 level compared to the J=3 level. Such a discrepancy would be further exacerbated if we used more accurate estimates of the dust temperature from SED modeling, which find that the dust is much warmer than our simple blackbody estimate \citep{li03}. The low gas temperature runs contrary to models of gas heating and cooling, which predict T$_{\rm gas}>$T$_{\rm dust}$ \citep{kam01}, although similar behavior is seen in the debris disk around HD 21997 \citep{kos13}.

\subsection{Residual Features}
Within the CO(1-0) data, the v=2.5km s$^{-1}$ channel is substantially brighter than the model (Figure~\ref{imspec},\ref{residuals_co10}) with a residual flux of 0.14 Jy/beam and a S/N of 4.7. It is offset from the phase center of $\Delta$RA=0.2$\arcsec$ and $\Delta$Dec=1.4$\arcsec$, is unresolved in the $\sim4\arcsec$ beam and no comparable feature is seen in the red-shifted side of the line, leading to a large asymmetry in the spectrum. This unresolved feature may be suggestive of an additional emission component, such as a planet or cloud of gas, within the disk. Its spatial offset corresponds to a distance of 160 au from the star, and its velocity is consistent with Keplerian motion at this distance, consistent with a body in orbit around the central star. Assuming a temperature of 25 K, as expected from our derived temperature structure at the observed distance of this feature, its flux implies a CO mass of $10^{-9}$ M$_{\odot}$, or a gas mass of $10^{-5}$ M$_{\odot}$ assuming CO/H$_2$=10$^{-4}$. While this feature is prominent in the CO(1-0) data, there are only hints of residual emission at a similar position in the CO(3-2) data, with a total flux of 0.11 Jy km s$^{-1}$ integrated over the four CO(3-2) channels that overlap with the single CO(1-0) channel. The low CO(3-2)/CO(1-0) flux ratio implies an excitation temperature of only 5 K. This cold, compact feature may be a result of asymmetries in the disk structure, similar to what has been seen in the CO gas of the debris disk around $\beta$ Pic \citep{den14}. Further data are needed to confirm the nature of this residual feature. 

In the CO(3-2) data there are some 3-5$\sigma$ residuals, corresponding to $\sim$10-30\%\ of the peak flux, that may reflect deviations from our simple model due to e.g. non-Keplerian motion or asymmetries in the disk. Higher spatial resolution ALMA observations of CO(3-2) also find evidence of asymmetric emission in the disk (White et al. in prep). While we assumed LTE in our modeling, the densities in our best fitting disk are starting to approach the critical densities of these lines (n$_{\rm crit}\sim10^{3}-10^{4}$ cm$^{-3}$), which may contribute to the imperfect model. To test our assumption of LTE at such low densities we input our density and temperature structure into the line radiative transfer code LIME \citep{bri10}, which is capable of performing the full NLTE calculation. NLTE and LTE spectra are indistinguishable indicating that NLTE effects are minimal and justifying our use of the LTE assumption. Since the residuals do not lead to large discrepancies between the model and data spectra outside of the bright CO(1-0) feature, and we can rule out strong NLTE effects, our results are likely representative of the true disk structure.

\section{Discussion}
The HD 141569 system has been referred to in the literature as both a transition disk \citep{esp14} and a debris disk \citep{zuc95}. Its young age, gas disk and active accretion are reminiscent of a protoplanetary disk, although the strength of the infrared excess is much more typical of a debris disk. With constraints on the temperature and density structure of the gas we can compare to typical protoplanetary disks and debris disks to gain insight as to the origin of the gas and dust in the system, as well as their influence on each other \citep[e.g.][]{wya15}.

Around HD 141569, we observe CO gas that extends out to $R_{\rm out}$=224$^{+21}_{-20}$ au with a surface density power law index of $\gamma=$1.7$^{+1.0}_{-1.1}$. Typical protoplanetary disks have outer radii that cover a wide range from 15-200 au \citep{hug08b,and09,and10,ise09} \footnote{Protoplanetary disks are often fit with an exponentially tapered power law instead of a straight power law. In these tapered models the characteristic radius, R$_c$ is {\it not} the same as the outer edge of a truncated power law, with typically R$_{\rm out}\sim$2R$_c$ for $\gamma\sim1$ \citep{hug08b}}. Within these tapered power law profiles, the power law portion has an average index of 1, although with very large dispersion, and much of the power law portion is unresolved \citep{and09,and10,ise09}. These protoplanetary disk profiles are similar to that of the HD 141569 system. 

The large inner radius is also similar to those seen in transition disks \citep{esp14} although in HD 141569 we find that the CO inner radius is similar to the inner radius of the mid-infrared dust emission \citep{fis00,mar02}, while in transition disks CO has been found to extend inside the dust gaps \citep{sal09,vanm15}. Active accretion, at a rate of $\sim$6$\times$10$^{-9}$ M$_{\odot}$ yr$^{-1}$ \citep{gar06,sal13}, indicates that gas exists within the CO inner radius although its exact spatial distribution is unknown. Gas that is devoid of CO may extend continuously from the CO inner radius toward the star; photodissociation of CO should occur in the area close to the star that is directly exposed to high energy radiation from HD 141569. 


While the spatial distribution is similar to protoplanetary disks, the mass of the HD 141569 disk is much smaller than in younger systems. Typical protoplanetary disks have gas masses of $\sim$10$^{-4}$-10$^{-2}$ M$_{\odot}$ \citep{and13,car14,wil13,man15} while we infer $M_{\rm gas}$=3.8$^{+15}_{-2.8}\times10^{-5}$ M$_{\odot}$ in the HD 141569 system. Conversely, \citet{pas06}, using mid-infrared spectroscopy and single disk mm CO spectroscopy, put upper limits on the gas in the outer radii of debris disks of a few earth masses, below what we derive in the HD 141569 system. Our gas mass is similar to that derived using CO emission from cold gas in the disks around 49 Ceti, with $M_{\rm gas}$=$3.9\pm0.9\times 10^{-5}$ M$_{\odot}$ \citep{hug08}, HD 21997, with $M_{\rm gas}$=8-20$\times 10^{-5}$ M$_{\odot}$ \citep{kos13}, and HD 131835 with $M_{\rm gas}$=1.3$\pm$0.6$\times 10^{-5}$ M$_{\odot}$ \citep{moo15}, all examples of massive CO disks. Our dust mass, estimated from the 870$\micron$ continuum, is 1.8$\times10^{-4}$ M$_{\odot}$=0.6 M$_{\oplus}$ similar to the $\sim0.1$ M$_{\oplus}$ dust masses seen in debris disks at ages of $\sim$10-100 Myr \citep{wya08} and implies an average gas/dust mass ratio of $\sim$10. 

The low gas mass may be a result of the multiple processes that contribute to the removal of CO, and gas in general, from the system.  At the current rate of accretion the observed gas would be removed in only $\sim$2500 years. Even if CO-poor gas extended down to 0.1 au, with the same surface density profile as inferred from the outer disk, accretion would still be able to deplete the disk in $\sim$5000 years. A photoevaporative flow can also remove gas from the disk on similar timescales \citep[and references therein]{ale14}. The CO itself can be selectively removed through its photodissociation by high energy photons. The observed CO mass would be destroyed in $\sim$200 years, using the photodissociation rates of \citet{vis09} and only considering the interstellar radiation field. This CO depletion could be partially offset by a source of replenishments within the disk, such as the collisions of comets and subsequent sublimation and release of trapped volatiles \citep{zuc12}. A total of 1.6$\times$10$^{19}$ kg of CO would need to be released each year, which corresponds to roughly 12000 Hale-Bopp size comets, assuming M$_{\rm Hale-Bopp}$=1.3$\times10^{16}$ kg and a 10\%\ CO content. \citet{jon07} find that the mass of the HD 141569 system is on the border of regime where self-shielding significantly diminishes the depletion of CO, although their disk mass is an order of magnitude larger than what we have derived here. The accretion and photoevaporative flows, as well as the photodissociation of CO, make it unlikely that the system is in steady state but instead it is likely in the process of transitioning from a gas-rich protoplanetary disk to a gas-poor debris disk \citep{wya15}.

Our finding of CO gas confined to radii smaller than that of the scattered light rings has important implications for the origin of the small dust grains. In a protoplanetary disk, much of which is shielded from the stellar radiation by the dust itself, the motion of dust grains depends on how well-coupled they are to the gas \citep{wei77}. Once the disk becomes optically thin, the grains experience an outward force due to radiation pressure, which can often overwhelm the gravitational pull of the central star. The ratio of radiation to gravitational force is characterized by the parameter $\beta$, which scales with the stellar luminosity and inversely with the grain size \citep{bur79}. Grains with $\beta>$0.5 ($\sim$sub-micron sizes) are quickly blown out of the system on dynamical timescales, while larger grains settle into orbits at slightly sub-Keplerian velocities. For the larger grains the presence of gas can further affect these orbits. If the radiation pressure influenced orbit is such that the grains move slower than the gas then they will experience a tail-wind which increases their angular momentum. These grains will move outward until they reach the outer edge of the gas disk. For HD 141569 this can occur for small-ish grains (tens of microns), while larger grains ($\sim$ hundreds of microns) settle into stable orbits within the gaseuous disk where radiation pressure and gas drag balance \citep{tak01}. Poynting-Robertson drag will drive grains with sizes of tens to hundreds of microns inwards towards the star \citep{li03}. The combined effect of these processes is a disk with a ring of dust at the outer edge of the gaseous disk, similar to what is seen in the HD 141569 system. 

In this scenario the small dust grains are generated in the inner disk and blown out on dynamical timescales. \citet{bil15} do not find any scattered light inside of 175 au, although they are limited by residuals from subtration of the central star within 0.5'' (=54 au). Resolved imaging in the mid-infrared has found small dust grains as close as 10-30 au \citep{fis00,mar02}. While our 870$\micron$ continuum observations are unresolved, at a spatial resolution corresponding to $\sim$150 au, higher resolution ALMA observations of the dust continuum find that the dust disk is compact and consistent with the mid-infrared observations (White et al. in prep). Collisions between planetesimals, either due to self-stirring  or through secular perturbations by a planet, could generate the inner dust seen in the mid-infrared which is then rapidly blown out of the system, populating the outer scattered light rings. Within the 5 Myr lifetime of the HD 141569 system, the Pluto-size planetesimals needed for self-stirring could be created as far out as $\sim$70 au \citep{ken08,ken10}, while a mildly eccentric planet (e$\sim$0.1) at the inner edge of the CO disk could perturb material out as far as 60 au \citep{mus09}. The relative youth of this system makes in-situ generation of small dust grains in the outer rings unlikley, unless by a planet embedded within the rings, but is consistent with their creation close to the star.




Generating small dust grains in the inner disk and having them get blown outward by radiation pressure is consistent with the relative location of the gas and dust, but it does not fully explain the morphology of the outer disk. \citet{tak01} predict a single ring, while two distinct rings are observed \citep{wei99,cla03,bil15}. A Saturn-mass planet embedded in the outer disk could carve out a gap and stir up the grains enough to create two rings \citep{wya05}, and \citet{bil15} find tentative evidence of a stellocentric offset that is consistent with perturbations from a planet. Another possibility relies on perturbations from one of the two distant companions \citep{wei00}. A close passage between one of the companions and the disk would truncate the disk, and excite dust grains just inside this truncation radius leading to the creation of small dust grains \citep{ard05,rec09}. Such a passage has been invoked to explain the large, low-surface brightness spiral features in the outer reaches of the disk \citep{cla03}.

Within the gas disk we derive an excitation temperature of $T_0$=27$^{+11}_{-4}$ K. Spitzer observations find typical dust temperatures of 50-200 K in debris disks \citep{mor11,bal13}, with similar results found by Herschel \citep{mat10,eir13,thu14}, although this depends strongly on the location and composition of the dust grains. In debris disks the dust temperature at a given radius is often found to be higher than that for blackbody grains, or conversely the observed radial location of the dust grains is larger than predicted for blackbody grains \citep{boo13}. Previous models of dust emission from the HD 141569 system \citep{li03,nil10} find typical dust temperatures of 50-200K in the outer disk, higher than our estimated blackbody temperature of 51 K, consistent with the presence of small dust grains. 

While in the HD 141569 system we find a gas temperature that is lower than that of the dust, it is important to keep in mind that we directly measure a gas excitation temperature that, under certain conditions, can be equated to the gas kinetic temperature and the dust temperature. \citet{kos13} find that in HD 21997 the gas excitation temperature implied by the CO lines is much lower than the expected dust temperature, similar to what is seen here. The deviation of the gas temperature from an optically thin disk of blackbody grains may be a property of a low density disk in which the collisions of the gas and dust are not frequent enough to equilibrate their temperatures. While the midplane gas densities reach 10$^{5}$-10$^{6}$ cm$^{-3}$, without detailed modeling of the dust emission we cannot know if the gas is well-mixed with the dust or if it is subject to a much different radial distribution. The low density environment in debris disks can lead to substantial differences between the gas and dust temperature \citep{kam01}, although these models predict the gas temperature to be much higher than the dust temperature. The HD 141569 disk is more massive than the $\beta$ Pic and Vega type disks considered by \citet{kam01}, which could lead to additional processes lowering the disk temperature. \citet{kam01} predict [OI] and [CII] play a large role in gas cooling, while \citet{thi14}, in their modeling of the HD 141569 system, find that CO itself can play a major role. Strong far-infrared emission from [OI] and [CII], at levels more comparable to protoplanetary disks than to debris disks, has been measured from the disk around HD 141569 \citep{mee12} although these lines may only contribute significantly to the upper low density regions of the disk rather than the midplane being traced by our CO observations. Two-dimensional photodissociation region models \citep[e.g.][]{hol99} predict that at the densities where CO becomes the dominant coolant, but before collisional equilibrium with the dust sets in, the gas temperature can drop below that of the dust. With dust temperatures of $\gtrsim$50K, CO will not freeze-out onto dust grains but instead will continue to exist in the gas phase where it can continue to efficiently radiate away thermal energy.

Comparing our derived gas temperature to these models assumes that the excitation temperature is equivalent to the gas kinetic temperature, while NLTE effects can cause these two temperatures to diverge. Our NLTE vs LTE test rules out significant NLTE effects, but it did so under the assumption of CO/H$_2$=10$^{-4}$. A diminished H$_2$ abundance will lead to fewer collisions and a deviation between the excitation temperature inferred by our data and the kinetic temperature predicted by the models. \citet{mat15} find that, in the case of the debris disk around Fomalhaut, the density and nature of the collision partners can induce NLTE effects that strongly affect the gas mass derived from CO upper limits. Further modeling is needed to fully explore the possibility of a low-density NLTE dominated disk as a fit to our observed CO emission.

\section{Conclusion}
The disk around HD 141569 likely represents a stepping stone between a fully optically thick primordial disk and a more evolved debris disk \citep{wya15}. Using CO(3-2) and CO(1-0) we have spatially resolved the cold gas emission from the disk around HD 141569. We have found that the gas has a large inner hole, and extends out to, but does not overlap with, the outer scattered light dust rings. This configuration is consistent with the generation of small grains within the inner disk, followed by interactions with the gas and the stellar radiation field that pushes them outward until they collect just beyond the outer edge of the gaseous disk. The dust may then be further stirred by a recent passage of one of the companions creating the observed multi-ring structure. While the CO gas mass is much lower than seen in protoplanetary disks, the active accretion and wide radial extent are consistent with a primordial protoplanetary disk origin, rather than the release of CO through collisions of gas-rich comets. We also find that the gas excitation temperature is colder than the equivalent blackbody dust temperature. This is consistent with the behavior seen in the gaseous debris disk HD 21997 \citep{kos13} but inconsistent with predictions from models \citep{kam01,jon07,thi14} possibly due to enhanced cooling from CO itself \citep[e.g.][]{hol99}, or NLTE effects \citep[e.g.][]{mat15}. Overall the data indicate a circumstellar disk in a later stage of its protoplanetary evolution, with a diminishing gaseous disk that may not be large enough to create a gas giant planet, but can still influence the distribution of the small dust grains.

\acknowledgements
We thank the referee for helpful comments regarding the context and interpretation of our results. We gratefully acknowledge support from NSF grant AST-1412647. The Submillimeter Array is a joint project between the Smithsonian Astrophysical Observatory and the Academia Sinica Institute of Astronomy and Astrophysics and is funded by the Smithsonian Institution and the Academia Sinica. The authors wish to recognize and acknowledge the very significant cultural role and reverance that the summit of Maunakea has always had within the indigenous Hawaiian community. We are most fortunate to have the opportunity to conduct observations from this mountain. This research made use of Astropy, a community-developed core Python package for Astronomy \citep{ast13}. Support for CARMA construction was derived from the states of California, Illinois, and Maryland, the James S. McDonnell Foundation, the Gordon and Betty Moore Foundation, the Kenneth T. and Eileen L. Norris Foundation, the University of Chicago, the Associates of the California Institute of Technology, and the National Science Foundation. CARMA development and operations were supported by the National Science Foundation under a cooperative agreement, and by the CARMA partner universities.

\clearpage

\end{document}